\documentclass[prr,superscriptaddress,amsmath,amssymb,twocolumn]{revtex4-2}
\usepackage{graphicx}
\usepackage{subfigure}
\usepackage{adjustbox}
\usepackage{bm}
\usepackage{bbm}
\usepackage{color}
\usepackage{braket}
\usepackage{standalone}
\usepackage{multirow}
\usepackage{tikz}
\usepackage{mathrsfs}
\usepackage{dsfont}
\usepackage[colorlinks,bookmarks=true,citecolor=blue,linkcolor=blue,urlcolor=blue]{hyperref}
\usepackage{cleveref}
\usepackage{comment}
\usepackage{mathtools}
\usepackage{soul}

\def\re{\text{Re}}

\def\tr{\text{Tr}}

 \Crefname{equation}{Eq.}{Eqs.}
\Crefname{figure}{Fig.}{Figs.}

\def\u{\underline}

\begin{document}
\title{Dissipative Boundary State Preparation}

\author{Fan Yang}
\affiliation{Institute of Physics, {\'E}cole Polytechnique F{\'e}d{\'e}rale de Lausanne (EPFL), CH-1015 Lausanne, Switzerland}
\ 
\author{Paolo Molignini}
\affiliation{Department of Physics, Stockholm University, AlbaNova University Center, 10691 Stockholm, Sweden}
\author{Emil J. Bergholtz}
\affiliation{Department of Physics, Stockholm University, AlbaNova University Center, 10691 Stockholm, Sweden}
\date{\today}

\begin{abstract}
We devise a generic and experimentally accessible recipe to prepare boundary states of topological or non-topological quantum systems through an interplay between coherent Hamiltonian dynamics and local dissipation. 
Intuitively, our recipe harnesses the spatial structure of boundary states which vanish on sublattices where losses are suitably engineered. This yields unique non-trivial steady states that populate the targeted boundary states with infinite life times while all other states are exponentially damped in time. 
Remarkably, applying loss only at one boundary can yield a unique steady state localized at the very same boundary.
We detail our construction and rigorously derive full Liouvillian spectra and dissipative gaps in the presence of a spectral mirror symmetry for a one-dimensional Su-Schrieffer-Heeger model and a two-dimensional Chern insulator. 
We outline how our recipe extends to generic non-interacting systems. 
\end{abstract}
\maketitle

\section{Introduction} Dissipation is ubiquitous and traditionally seen as detrimental to quantum phenomena. 
Much effort is thus aimed at minimizing its effects. 
Recently, however, it has been realized that \emph{structured} dissipation can instead lead to new and intriguing topological physics \cite{diehl2011,Gong18,Bergholtz21,Miri2019,Ashida2020}. 
In the quantum realm, an example thereof are dynamical Liouvillian skin effects \cite{fei2019, ueda2021, emil2022l,kohei2023} reflecting the underlying non-Hermitian (NH) topology \cite{Lee2016,yao2018,flore2018,lin2023,okuma2023} on the level of quantum master equations.
Another exciting aspect is the preparation of topological steady states of matter through judiciously engineered dissipation, which provide an alternative to ground state cooling \cite{diehl2011,bardyn2013,budich2015,Goldman19,Goldstein2019,Shavit2020,Tonielli2020,Bandyopadhyay2020,liu2021,jens2022}. 
A preeminent idea considers the limit of strong dissipation in which jump operators \cite{lindblad1976} mimic localized Wannier functions and facilitate a projection onto the pertinent (`low energy') subspace \cite{diehl2011,bardyn2013,budich2015}. 
Yet, while conceptually appealing, such approaches are both extremely challenging to implement and face a number of fundamental obstructions \cite{budich2015,Goldman19,Goldstein2019}.

\begin{figure}[t]
\centering
\includegraphics[width=\columnwidth]{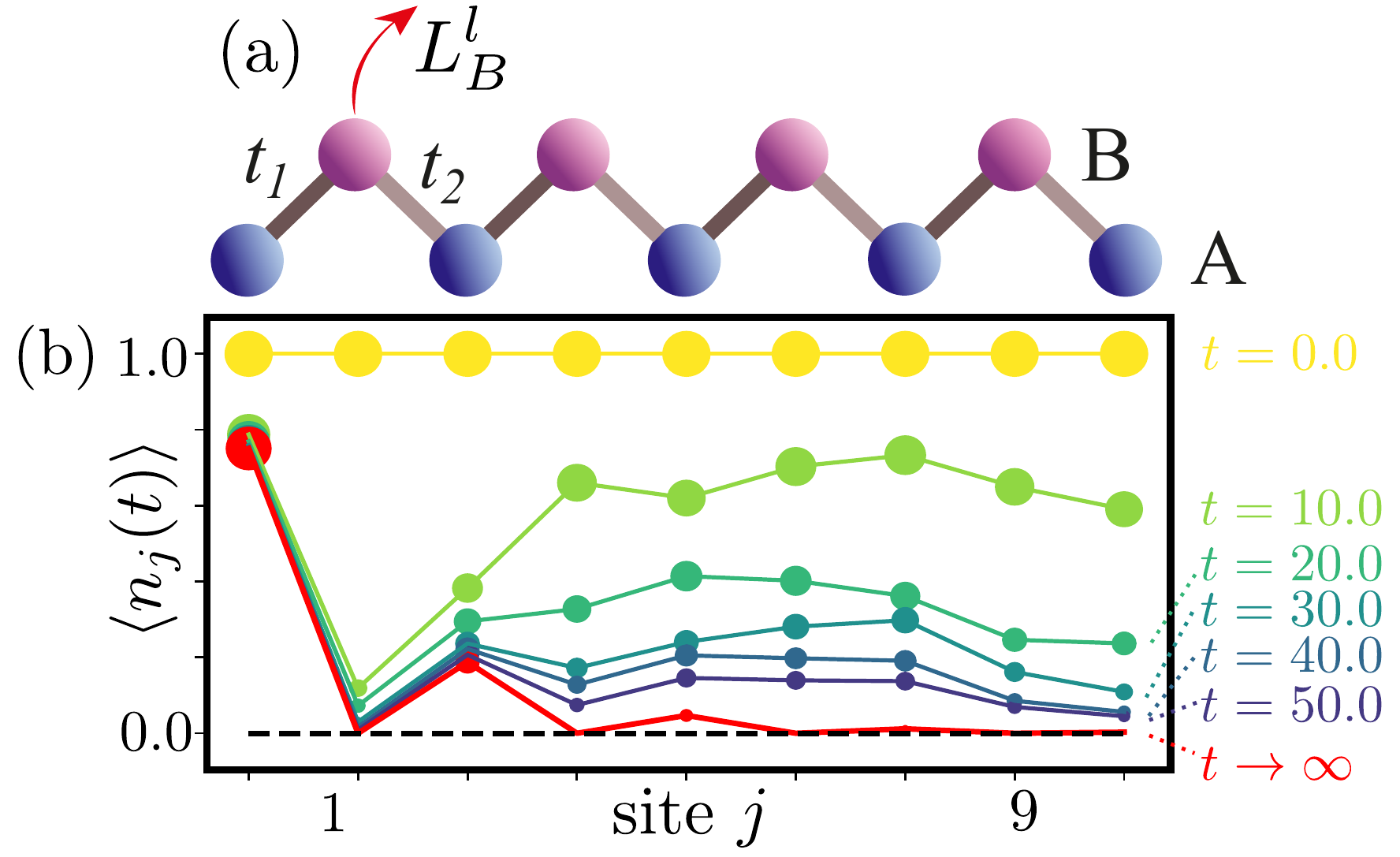}
\caption{{\bf Minimal example: boundary state from single site loss}. An open SSH chain with local loss only on the first site of the B sublattice (a), exhibiting a dynamical damping towards the boundary state at the same boundary (b) as indicated by the particle density $\left< n_j(t) \right>$ for a chain of length $L=2N-1=9$ with $t_1 = 0.5$, $t_2=1.0$ and $\gamma_B=0.5$.
}
\label{fig:sketch}
\end{figure}

Here, we consider an alternative approach that crucially depends on the interplay of both coherent dynamics and dissipation. 
This approach alleviates both practical and fundamental challenges: with directly accessible ingredients, we can devise a combination of Hamiltonian dynamics and dissipation whose interplay uniquely prepares the system in what is the key hallmark of a topological phase, namely its boundary states. 
At long times all bulk modes vanish and only the boundary states prevail. 
A minimal example is shown in Fig. \ref{fig:sketch}. 
Adding loss on a \emph{single} site in a Su-Schrieffer-Heeger (SSH) chain evolves the system into a state localized at the same boundary. 
Our recipe is generic and can be applied to essentially any topological (or non-topological) noninteracting system. 
It also does not rely on fine tuning. 
Instead, the success of the construction is rooted in choosing lattices harboring boundary modes with the unique property of being the only eigenstates that vanish exactly on some sublattices \cite{flore2017,flore2018l,flore2019b,flore2019e}---where we engineer loss. 
In fact, in the SSH example of Fig. \ref{fig:sketch}, the same boundary steady state is obtained whenever loss is applied to one or more of the B-sites. 

\begin{figure}[t]
\centering
\includegraphics[width=\columnwidth]{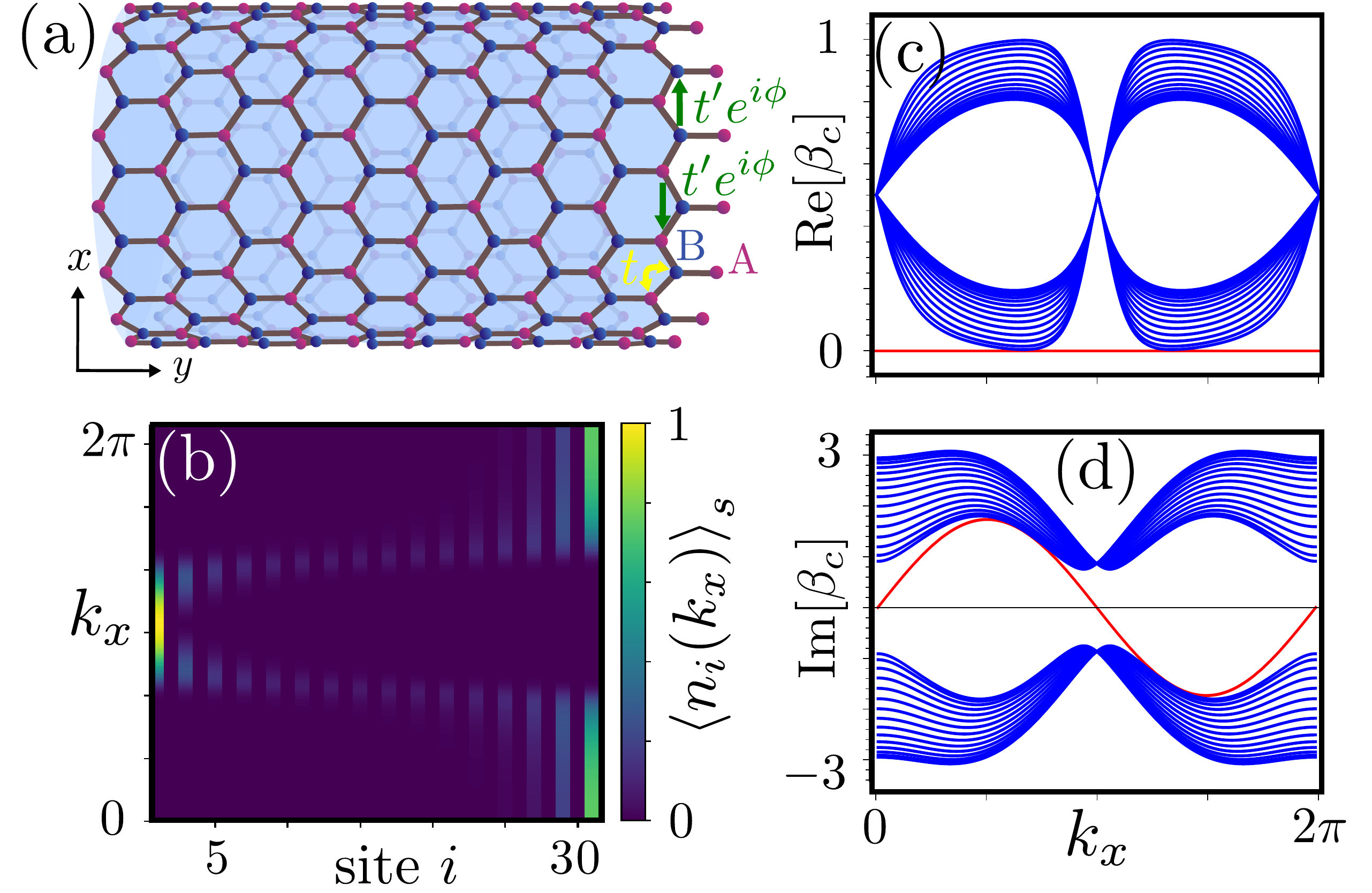}
\caption{{\bf Preparation of chiral Chern insulator edge states}. (a) Chern insulator on a cylinder geometry with (b) the distribution of the steady chiral mode prepared by B-sublattice loss as a function of $k_x$ along the $y$ direction. We adopt $L_y=2N-1=31$ for OBC and $L_x=200$ for PBC. The chiral edge state is localized at the left (right) boundary when $|r(k_x)|=|2\cos(k_x/2)| < 1$ ($>1$). The localization position changes at $k^*_{x} = 2\pi/3, 4\pi/3$ when 
$|r(k^*_x)|=1$. 
(c),(d) Complex rapidity spectrum of the Chern insulator for $N=16$ and $t'/t=\sqrt{3}/2$, $t=\gamma=1$ demonstrating bulk modes (blue lines) and exact chiral mode (red line). The real part (c) illustrates the dissipative gap and the imaginary part (d) is equivalent to the full energy spectrum.}
\label{fig:chern}
\end{figure}

We derive the aforementioned results analytically, and in the presence of a spectral mirror symmetry we obtain the full exact Liouvillian spectrum of the corresponding Lindblad master equation. We exemplify our construction for SSH chains with even and odd number of sites (Figs.~\ref{fig:sketch}-\ref{fig:scaling}) and for a two-dimensional (2D) Chern insulator hosting exact chiral edge steady states (Fig. \ref{fig:chern}). 
We also consider the effect of gain yielding steady state currents and how our exact solutions provide key intuitions  for non-solvable systems (cf. Fig. \ref{fig:scaling}).

\section{ Setup and dissipative SSH models} We consider the Lindblad master equation \cite{breuer2007, lindblad1976} which {captures} quantum dynamics of Markovian dissipative systems \cite{diehl2008,kraus2008,verstraete2009,krauter2011,langen2015},
   \begin{gather}
     \frac{d \rho}{dt} = \hat{\mathcal{L}} \rho \coloneqq  -i [ \mathcal{H}, \rho ] + \sum_{\mu} ( \hat{L}_\mu \rho \hat{L}_\mu^\dagger
     - \frac{1}{2}\{ \hat{L}_\mu^\dagger \hat{L}_\mu, \rho \} ),
      \label{eq:lin} 
\end{gather}
where $\rho$ is the density matrix of the system and $\mu$ denotes the summation over all types of jump operators. 

As a minimal example, we study an SSH model of spinless fermions with an odd number of sites $L= 2N - 1$, with Hamiltonian
$\mathcal{H}_{S} = \sum_{j=1}^{N-1} t_1 a_{j,A}^\dagger  a_{j,B}+t_2 a_{j+1,A}^\dagger  a_{j,B}+ \text{H.c}$.
We add local loss on the first {B site}: $\hat{L}^l_{1,B}=\sqrt{\gamma^l_{0,B}} a_{1,B}$ [see Fig.~\ref{fig:sketch}(a)].
Here, $a_{j,A}^\dagger (a_{j,B})$ are creation (annihilation) operators on the sublattice A(B) in the $j$th unit cell satisfying anticommutation relations: $\{a_{j,\alpha}, a_{j',\alpha'}^\dagger\}=\delta_{j,j'} \delta_{\alpha,\alpha'}$. 
The quadratic Hamiltonian and linear Lindblad {dissipators} yield a quadratic Lindbladian {diagonalizable} in the third quantization approach \cite{prosen2008, prosen2010ex, prosen2010sp}. 
The damping matrix $X$ which governs the dynamics can be mapped to a {NH} tight-binding Hamiltonian encoding information of both the original Hamiltonian and Lindblad dissipators (see \Cref{app:liou}),
\begin{gather}
 X = \frac{\gamma}{2}\times I_{L \times L} + iU \mathcal{H}_{\text{NH}}U^{-1}, \ 
 \mathcal{H}_{\text{NH}} = \mathcal{H}_S+ i \Upsilon,
 \label{eq:x}
 \end{gather}
{under a} unitary transformation $U = \text{diag}\{1,i,1,i\dots, 1\}$. For a single loss on the first $B$ site, $\gamma = \gamma_B = |\gamma_{0,B}^l|/2$, $I = \text{diag}\{1,1,0,\dots\}$ and $\Upsilon = \text{diag} \{\gamma/2, -\gamma/2, 0,\dots,0 \}$. 
The damping matrix $X = X_{c(d)}$ denotes the equal contribution from two Majorana fermions species: $a_{j,A}=\frac{1}{2}(c_{j,A} - id_{j,A}), a_{j,B}=\frac{1}{2}(d_{j,B} + ic_{j,B})$. 
It can also be decomposed into its eigenmodes: $X = \sum_m \beta_m |\psi_{Rm}\rangle \langle \psi_{Lm}|$ with $m$ the band index. 
The left and right eigenvectors satisfy the biorthogonal relations \cite{brody2013, flore2018, elisabet2020}: $\u{\psi}^*_{L, m} \cdot  \u{\psi}_{R, m'} = \delta_{m,m'}$. 

This decomposition enables us to derive dynamical observables by integrating the Lindblad master equation.
The particle number $\langle n_j(t)\rangle = \tr[a_j^\dagger a_j\rho(t)]$ reads (see \Cref{app:liou})
 \begin{align}
 & \langle n_j(t)\rangle - \langle n_{j} \rangle_s  
 \label{eq:nm}   \\
    &= \sum_{m,m'} \sum_{l =1}^{L} e^{-(\beta_m + \beta^*_{m'})t} \psi_{Lm'} (j)  \psi^*_{Rm'} (l) \psi_{Rm} (l) \psi_{Lm}^* (j),\notag
\end{align}
where $\langle n_{j} \rangle_s = \langle n_j(t = \infty)\rangle = 0$ corresponds to the trivial non-equilibrium steady state (NESS) in the presence of loss, i.e. an empty chain, and the initial condition is chosen as a completely filled chain.
The eigenvalues $\beta_m$ of the damping matrix coincide with the rapidity spectrum of the Liouvillian \cite{prosen2008,emil2022l}.
Its real part encodes the {decay rates} of different modes {towards} the NESS {and we define} the dissipative gap $\Delta = 2\min \{ \re[\beta_m] \} \ge 0$.

For an odd number of sites, $\mathcal{H}_S$ hosts a zero-energy boundary mode fully suppressed on the $B$ sublattice~\cite{flore2019e}: $\mathcal{H}_S \u{\psi}_0 = E_0\u{\psi}_0$, 
$E_0 = 0$ and  
\begin{gather}
 \u{\psi}_0 = \mathcal{N}(r,0,r^2,0,\dots, 0, r^N)^T,
 \label{eq:bm}
 \end{gather} 
{with} $r=-t_1/t_2$ the localization factor and $\mathcal{N}^2=(1-r^2)/[r^2(1-r^{2N})]$ a normalization. 
For $|r| < 1$  $( > 1)$, the boundary mode is exponentially localized at the left (right) end of the chain. 
Intuitively, the frustrated nature of the boundary mode indicates {its robustness} against local loss on any B site. 
Indeed, with a single loss on the first $B$ site, we find a vanishing rapidity for this boundary mode: 
$X \u{\psi}_{R0} = \beta_0 \u{\psi}_{R0}$,  $X^\dagger \u{\psi}_{L0} = \beta^*_0 \u{\psi}_{L0}$ where $\beta_0 = 0$ and $\u{\psi}_{R0} = \u{\psi}_{L0} = \u{\psi}_0 $ while all bulk modes have a finite dissipative gap. 
It implies that starting from all sites that are filled, through the dissipation the system always selects the boundary mode as the non-trivial steady state. 
For $t \to \infty$, the particle number becomes
\begin{equation}
 \langle n_j \rangle_{s}=\frac{r^{j-1}-r^{j+1}}{1-r^{2N}} \:\: (j \: \mathrm{odd}), \quad \langle n_j \rangle_{s}=0 \:\:  (j \: \mathrm{even}).
 \label{eq:ns}
\end{equation}

%
%
To verify our observation, 
in Fig.~\ref{fig:sketch}(b), we calculate the time evolution of the particle number on individual sites by numerically diagonalizing the damping matrix. 
At sufficiently long times, the boundary mode (red line), with  $|r|=|-t_1/t_2| = 0.5$ and site occupation $\langle n_j \rangle_{s}$ predicted by Eq.~(\ref{eq:ns}), exponentially localizes ($|r| < 1$) even when the single B-site loss is placed close to the same left end of the chain.
We also observe that with single weak loss ($\gamma/2 \le \left||t_1|-|t_2|\right|$), the bulk dissipative gap is inversely proportional to the chain length: $\Delta_{\text{bulk}} \propto \gamma/N$. 
The single loss of our minimal model is particularly useful for small system size in modern experimental setups.

\begin{figure}[t]
\centering
\includegraphics[width=\columnwidth]{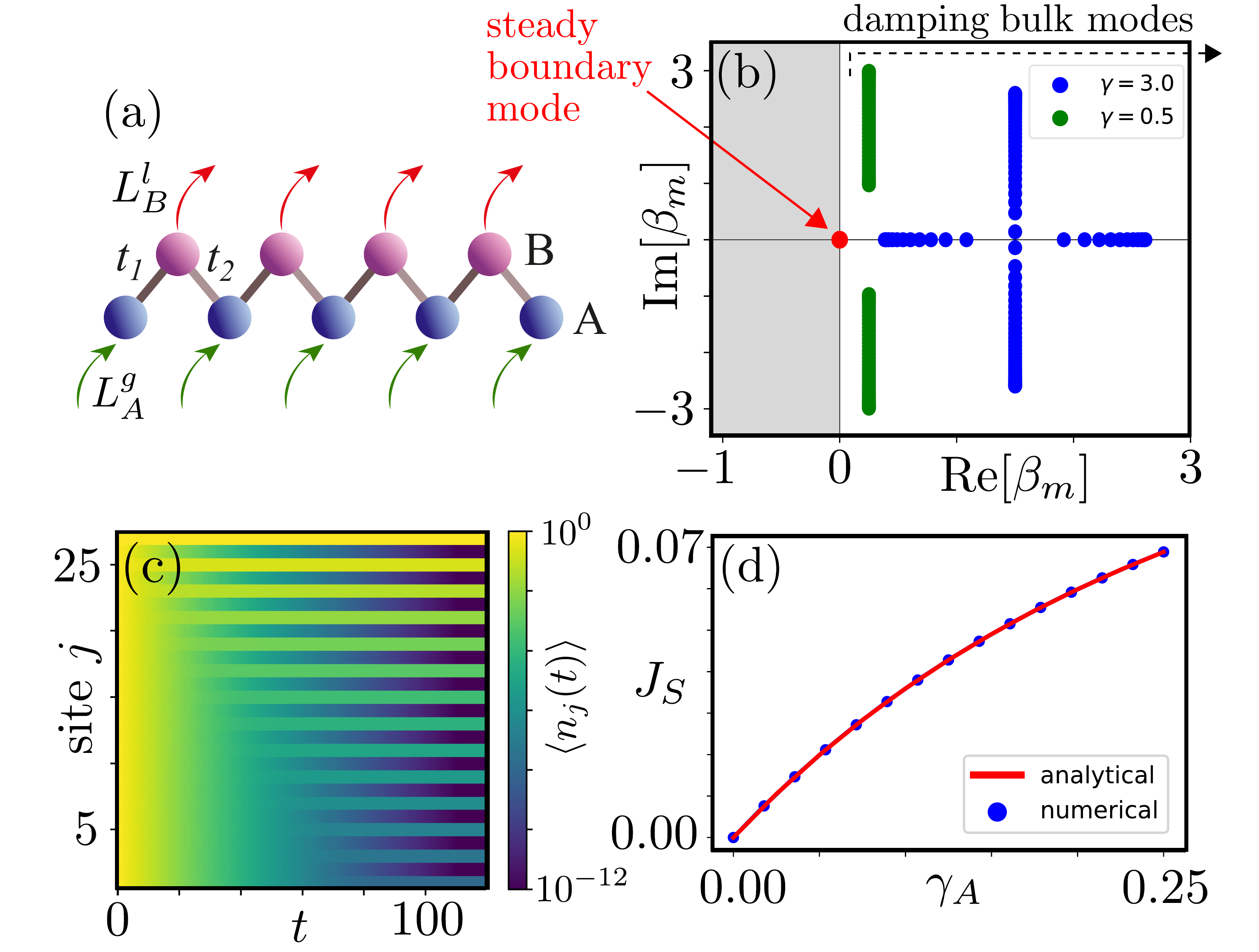}
\caption{{\bf Solvable SSH chain with odd number of sites and uniform gain and loss}.
(a) Illustration of an open SSH chain with loss rate $\gamma_B$ on the B sublattice and gain rate $\gamma_A$ on the A sublattice.
(b) Rapidity spectrum for $\gamma_B = 0.5$ (green dots) and $\gamma_B= 3.0$ (blue dots). The red dot at zero corresponds to the steady boundary mode.
(c) Dynamics of the particle density  $\left< n_j(t) \right>$ for a chain of length $L=2N-1=27$ with $t_1=2.0$, $t_2=1.0$, $\gamma_A=0.0$, $\gamma_B=0.5$.
(d) Steady-state current $J_S$ as a function of $\gamma_A$. The other parameters are as in (c).}
\label{fig:ssh-odd}
\end{figure}

In the second example, while keeping the dissipationless feature of the boundary mode, we can make the dissipative gap of the bulk modes saturate in the large system size limit by adding {loss on the entire B sublattice:} \textcolor{black}{$\hat{L}^l_{j,B}=\sqrt{\gamma^l_{0,B}} a_{j,B}, \forall j$}  [red arrows in Fig.~\ref{fig:ssh-odd} (a)]. 
The damping matrix $X$ in Eq.~(\ref{eq:x}) holds new entries with $I = \mathbbm{1}_{L \times L}$ and $\Upsilon = (\gamma/2) \times \text{diag} \{1, -1, 1,\dots,-1, 1 \}$. The rapidity for the boundary mode in Eq.~(\ref{eq:bm}) remains rigorously zero, $\beta_0 = 0$, and it is thus selected as the non-trivial steady state again.
This steady boundary mode with a vanishing Liouvillian gap [denoted by the red dot in Fig.~\ref{fig:ssh-odd}(b)] is not included in Ref.~\cite{cooper2020}, while it is referred to as an `{edge} dark state' in Ref.~\cite{fei2023}. 
Remarkably, we can obtain the {bulk dissipative gap} analytically by exploiting spectral {mirror symmetry under open boundary condition (OBC) with odd sites}.
The rapidity spectrum of the $(2N-2)$ bulk modes comes in pairs $\beta_m = \gamma/2 \pm iE_m$, where $E_m$ {denote} the eigenenergies of $\mathcal{H}_{\text{NH}}$ from the mapping of Eq.~(\ref{eq:x}) and we have adapted the notation for the band index $m=(\pm, q)$ with $q = \pi m'/N$, $m'=1,2,\dots, N-1 $. 
The spectrum obeys {spectral} mirror symmetry $\beta^{{\text{OBC}}}_\pm (q) = \beta^{{\text{OBC}}}_\pm (-q)$, establishing a direct link to the spectrum under periodic boundary condition (PBC) \cite{flore2019e,elisabet2020,elisabet2022}: $\beta^{\text{OBC}}_\pm (q) =\beta^{\text{PBC}}_\pm (q)$, reflecting the absence of a {NH} skin effect. We thus obtain $\beta_{\pm}^{\text{OBC}}(q) = \frac{\gamma}{2} \pm i \sqrt{t_1^2 + t_2^2 + 2t_1t_2 \cos(q) - \frac{\gamma^2}{4}}$. 
{Shown} in Fig.~\ref{fig:ssh-odd}(b), all bulk modes have a finite gap and the minimum  $\Delta_{\text{bulk}} = 2\min \{ \re[\beta_\pm (q)] \}$ determines {the dissipative preparation time for the steady boundary mode: $\tau \sim \Delta_{\text{bulk}}^{-1}$.}
In the large-$N$ limit, we find the analytical solution to the {bulk} dissipative gap: $\Delta_{\text{bulk}}=\gamma$, for $\left||t_1| - |t_2|\right| \ge {\gamma}/2$; $\Delta_{\text{bulk}}=\gamma - \sqrt{\gamma^2 - 4 (|t_1| - |t_2|)^2}$, otherwise.
An exact complete set of right and left eigenvectors for the damping matrix $X$ {is obtained as well} (see \Cref{app:liou}), where the bulk modes can be viewed as a superposition of two Bloch waves with opposite momenta vanishing on the last B site (which is removed in the odd chain). We are able to analytically resolve the full time evolution of the particle number shown in Fig.~\ref{fig:ssh-odd}(c).
For {weak dissipation with} $\Delta_{\text{bulk}} = \gamma$, the damping wavefront arises from the steady boundary mode, in contrast to previous studies with a major contribution from  bulk {skin} modes \cite{fei2019, ueda2021, emil2022l,kohei2023} (here our bulk modes are delocalized).

Next, we study the effect of small gain $\hat{L}^g_{j,A}=\sqrt{\gamma^g_{0,A}} a^\dagger_{j,A}, \forall j$ on the A sublattice [green arrows in Fig.~\ref{fig:ssh-odd}(a)]. 
The damping matrix $X$ in Eq.~(\ref{eq:x}) is still exactly solvable  with entries $\gamma = \gamma_A + \gamma_B$, $I=\mathbbm{1}_{L \times L}$, $\Upsilon = [(\gamma_B-\gamma_A)/2] \times \text{diag} \{ 1, -1, 1, -1, \dots, 1 \}$ {and} $\gamma_A = |\gamma^g_{0,A}|/2$.
The boundary mode remains an eigenmode, yet it is associated with a non-zero rapidity $\beta_0 = \gamma_A$ leading to a finite gap: $\Delta_0 = 2\gamma_A$. 
A small gain also eliminates the empty state as trivial NESS. 
In \Cref{app:ssc}, we identify the analytical structure of the non-trivial NESS in the presence of $\gamma_A$ as a mixed state continuously connected to the empty state and the boundary mode {when} $\gamma_A \to 0$.  
The system exhibits a steady state current $J_S = (i/L) \sum_j (\langle a_j^\dagger a_{j+1} \rangle_s - \langle a_{j+1}^\dagger a_j \rangle_s )$, which we can obtain in a closed form [see \Cref{eq:jss}].
We compare the analytical result with numerics in Fig.~\ref{fig:ssh-odd}(d). 
Once $0 < \gamma_A \ll \gamma_B$, regardless of the initial conditions, the system eventually evolves to the non-trivial NESS with a localization structure approximating the boundary mode in Eq.~(\ref{eq:bm}).

\begin{figure}[t]
\centering
\includegraphics[width=0.98\columnwidth]{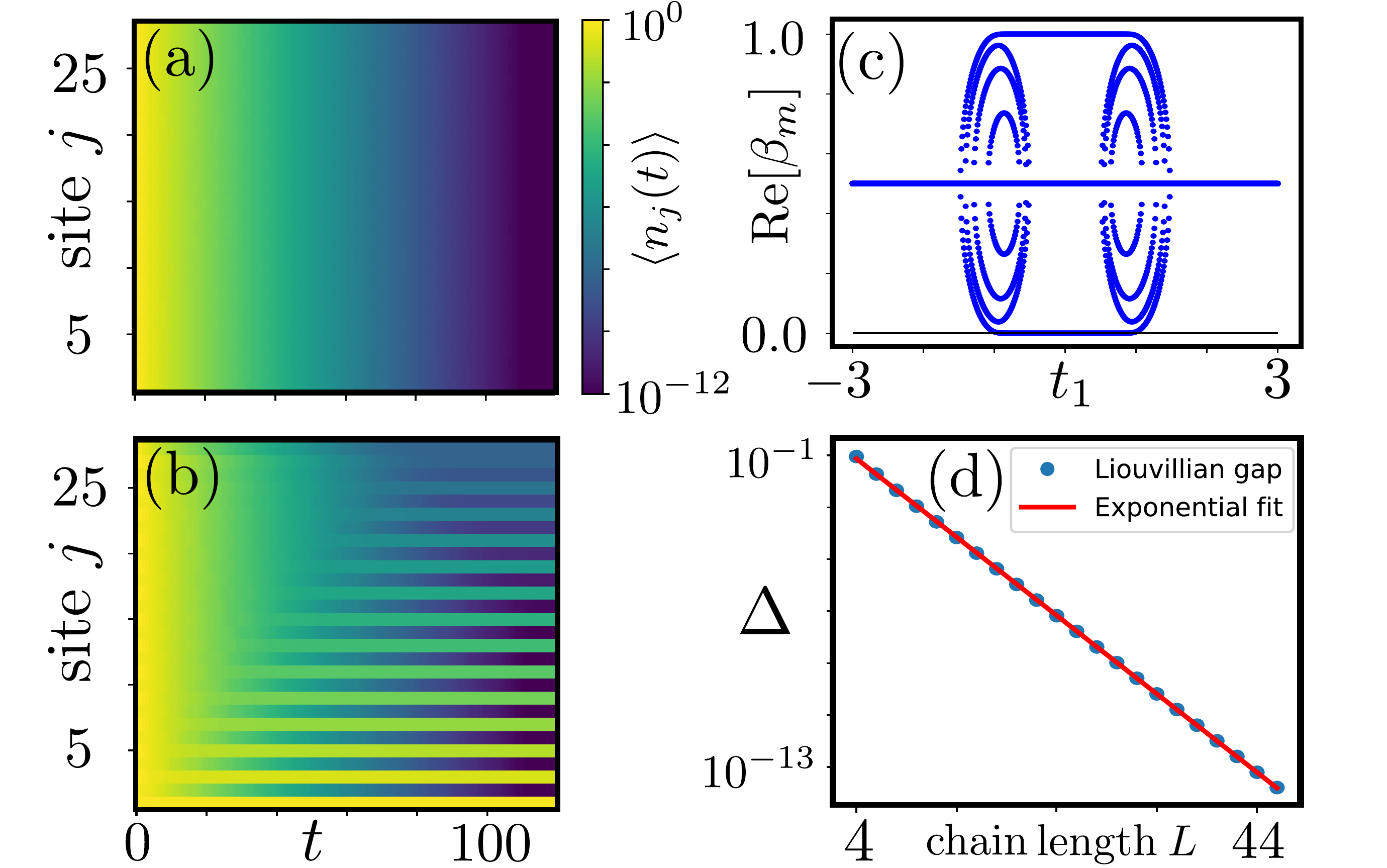}
\caption{{\bf Topology and scaling in SSH chains with even number of sites and uniform loss}.
(a) Dynamics of the density $\left< n_j(t) \right>$ for a topologically trivial chain of length $L=2N = 28$ with $t_1=2.0$, $t_2=1.0$, $\gamma_B=0.5$.
(b) Dynamics of the density $\left< n_j(t) \right>$ for a topological chain of length $L= 28$ with $t_1=0.5$, $t_2=1.0$, $\gamma_B=0.5$.
(c) Real part of the rapidity spectrum as a function of $t_1$ for a system with $L=46$ sites and $t_2=1.0$, $\gamma_B=1.0$.
(d) Dependence of the Liouvillian gap $\Delta = 2\min \{\re[\beta_m]\}$ on the chain length $L$ for an even number of sites. 
Blue dots are the numerical result, while the red line indicates an exponential fit $f(N) = C \exp(-\alpha N)$ with $C=1.202$ and $\alpha=0.695$. 
}
\label{fig:scaling}
\end{figure}

We now address the non-solvable model by obtaining the eigenmodes of the damping matrix through exact diagonalization (ED). 
The simplest scenario is encountered when we consider the SSH chain with an even number of sites $L=2N$.
For $L=2N-1$, the model is symmetric under the exchange of $t_1$ and $t_2$ bonds [see Fig.~\ref{fig:ssh-odd}(a)]. 
The boundary mode remains the steady state in both the topological ($|t_1| < |t_2|$) and non-topological ($|t_1| > |t_2|$) regions, with the localization direction reversed under the exchange ($|r| \to |r|^{-1}$).
For $L=2N$, topology comes into play.
With loss on the B sublattice, the boundary mode disappears in the topologically trivial region ($|t_1| > |t_2|$), as seen by comparing Fig.~\ref{fig:scaling}(a) with Fig.~\ref{fig:ssh-odd}(c). 
Yet, it is retrieved in the topological region ($|t_1| < |t_2|$) [see Fig.~\ref{fig:scaling}(b)] with the same localization factor $|r|=|t_1/t_2| < 1$ as in \Cref{eq:bm}. 
This phenomenon can be predicted by noticing that two zero-energy boundary modes of the SSH model {of even lengths} are localized at different ends, and only one of them is protected under the B-sublattice loss \cite{emil2023n,smith2023}. 
From the rapidity spectrum in Fig.~\ref{fig:scaling}(c), we observe this boundary mode only at $|t_1| < |t_2|$ and find that {its} dissipative gap decays exponentially with the chain length [see Fig.~\ref{fig:scaling}(d)]. 
{The} lifetime {of the topological boundary state} is enhanced exponentially by increasing the system size: $\tau \sim \Delta^{-1} \sim \exp{(\alpha N)}$ with $\alpha>0$, a remarkable feature due to topological protection.

\section{ Generalization to dissipative lattice models in any dimension}
Our recipe for devising unique steady boundary states naturally generalizes to any dimension with localization on boundaries of any co-dimension including surfaces, corners, edges and hinges. 
This is achieved by inferring results on boundary state solutions that vanish exactly on certain sublattices \cite{flore2018l,flore2019b}. In presence of a spectral mirror symmetry the full Liouvillian spectrum may be obtained \cite{flore2019e}. 
Increasing both the bulk dimension and the codimension simultaneously is straightforward, yielding e.g. corner steady states on the breathing kagome lattice \cite{Ezawa2018,flore2018l}, in full analogy with the SSH chain. 
Increasing the bulk dimension is suitably done by dimensional extension which is slightly more technically involved.
Here, we provide an explicit example of dimensional extension in the preparation of dissipationless chiral edge modes of a two-dimensional Chern insulator on the honeycomb lattice \cite{haldane1988, flore2019e}. 

Figure \ref{fig:chern}(a) illustrates the corresponding cylinder geometry: { we impose} PBC on the $x$ direction with an even number of sites $L_x$ and OBC along the $y$ direction with an odd number of sites $L_y = 2N-1$.
The real nearest-neighbor and next-nearest-neighbor hopping amplitudes are denoted as $t$ and $t'$. 
The complex hoppings $t'e^{i\phi}$ with $\phi=\pi/2$ are allowed only {between} unit cells along the {$x$} direction. 
For each $k_x$, a mapping to the generic SSH-like model can be established (see \Cref{app:chern}): $\mathcal{H}_{S} (k_x)=\sum_{j} \sum_{\alpha=A,B}  \epsilon_{\alpha} a_{j,\alpha}^\dagger a_{j,\alpha} + \sum_{j=1}^{N-1} t_1 a_{j,A}^\dagger  a_{j,B}+t_2 a_{j+1,A}^\dagger  a_{j,B}+ \text{H.c.}$, with $t_1=2t\cos(k_x/2), t_2=t,\epsilon_A=-\epsilon_B=-2t'\sin(k_x)$ (we use renormalized $k_x = \sqrt{3}k_{x,0}$ compared with the actual value).
With {the $B$-sublattice loss} $\gamma=\gamma_B$, the rapidity spectrum consists of two copies ($\mu=\pm 1$) for distinct Majorana fermions species ($c$, $d$). 
For the {topological} chiral mode, $\beta_0^\mu=-i\mu\epsilon_A$ while for the bulk modes, the spectral mirror symmetry $\beta_{\pm}^{{\mu}} (k_x,k_y) = \beta_{\pm}^{{\mu}} (k_x,-k_y)$ leads to $\beta_{{\mu},\pm}^{\text{OBC}} (k_x, k_y) = \beta_{{\mu},\pm}^{\text{PBC}} (k_x, k_y) = {\gamma}/{2}\pm i \sqrt{t_1^2+t_2^2+2t_1t_2\cos(k_y)+(i\gamma/2-\mu \epsilon_A)^2}$ {with} $k_y=\pi j/N$, $j=1,\dots,N-1$.
Figures \ref{fig:chern}(c) and \ref{fig:chern}(d) show the real and imaginary parts of the rapidity spectrum of the chiral mode (red) and bulk modes (blue). 
The bulk modes have a vanishing dissipative gap in the large system limit at the momenta $k^*_x$ which satisfies $|r(k^*_x)|=1$ where the chiral mode becomes delocalized and switches sides. Nevertheless for momenta when the chiral mode is isolated within the real rapidity spectrum of Fig.~\ref{fig:chern}(c), the system has a finite instantaneous $\Delta_{\text{bulk}}$ to separate the chiral and bulk modes for any system size, and thus inherits the localization structure to the boundary mode of the SSH model with $|r(k_x)|=|t_1/t_2|=|2\cos(k_x/2)|$. This feature is illustrated by the steady state particle number of Eq.~(\ref{eq:ns}) in Fig.~\ref{fig:chern}(b).

\section{Discussion} {In this work, we have explored an alternative to ground state cooling and Floquet engineering of topological phenomena \cite{Eckardt17,Cooper19,Goldman19,Aidelsburger2013,Jotzu2014,Molignini2023}, as well as other finite-time methods to probe boundary states \cite{goldman2013,leder2016}. Our generic approach utilizes \emph{both} dissipation and coherent dynamics to prepare boundary states as steady states which can be operated at arbitrarily long times. It circumvents the problem of thermal excitations faced at finite temperature in existing platforms. Compared with other dissipative preparation schemes \cite{diehl2011, bardyn2013}, in our recipe coherent dynamics from the Hamiltonian itself is introduced, while the form of dissipation is greatly simplified.} 

{Our} exact construction is not only simple, but also very general. By choosing appropriate lattice geometries it carries over directly to a plethora of topological and non-topological boundary states, including Fermi arcs and higher-order states at corners and hinges which all have the desired nodal boundary state structure \cite{flore2017,flore2018l,flore2019b,flore2019e}. {Going beyond the idealized model, one can break spectral mirror symmetry in the SSH Hamiltonians by adding disorder on the nearest-neighbor hopping terms. We find that the unique localization structure of the boundary mode is a generic property of any odd length chain: it still vanishes on the entire B sublattice without its dissipative gap opening, and is robust against onsite-potential disorder on any B site (see \Cref{app:dis}). In real experiments, weak disorder on the other sublattice and further-neighour interactions directly interfere with the boundary mode and should be precisely controlled. The latter can be effectively suppressed  on sufficiently deep optical lattices \cite{jaksch1998} that make the SSH models applicable \cite{Atala2013,atala2014,meier2016,leder2016}.}

{From the theoretical side, our analytical study offers promising outlooks for future exploration as well. One could test targeted boundary state distillation in open systems inherent with Liouvillian skin effects \cite{fei2019, ueda2021, emil2022l,kohei2023}, critical phenomena \cite{berdanier2019, zhang2022} and phase transitions \cite{nava2023,starchl2022,starchl2023}. In particular, the preparation of steady boundary modes with NH skin effects would bestow exponentially enhanced sensitivity on quantum sensing devices \cite{NTOS,clerk2020,QNTOS,parto2023,konye2023}. It is equally exciting to extend the current framework from open fermions to bosonic and even hybrid systems via a similar third quantization approach \cite{prosen2010q,barthel2022,medic2023}. }

Very recently, experiments on light in lossy optical waveguides showed that a similar boundary state preparation is possible at the level of NH effective Hamiltonians in classical settings \cite{emil2023n}. The present work shows that a similar NH phenomenology is also relevant in the quantum realm and opens up different avenues for quantum control. 


We also note that related systems with staggered loss or weak measurements have been considered earlier in the context of quantum walks described by effective NH Hamiltonians \cite{Rudner09,Rakovszky17,Zeuner2015}. There, however, rather than the preparation of topological boundary states the focus was on other aspects such as defects, phase transitions and mean displacement.

{Recently, we became aware of two independent works numerically observing the amplification of states at one boundary in Chern insulator models with gain and loss \cite{fei2023,meng2023}. 
In these studies, the boundary modes are classified and dynamically selected~\cite{emil2023n,Hyart2019,smith2023} as long-lived modes on even-length lattice models, but not as steady states once the system size becomes finite. By contrast, 
our predominantly analytical work focuses on odd-length lattice models and achieves this goal with generic boundary states hosting an exact zero dissipative gap.}  We also consider systems in which the entire family of boundary states is prepared, such as the Chern insulator model. 
This indicates a great flexibility and generality in harnessing structured dissipation for probing topological boundary state physics through its interplay with coherent Hamiltonian dynamics.

	\textit{Acknowledgements.--} We thank Mohamed Bourennane, Johan Carlström, Sebastian Diehl, and Kohei Kawabata for discussions.
This work was supported by the Swedish Research Council (Grant No. 2018-00313), the Knut and Alice Wallenberg Foundation (KAW) via the Wallenberg Academy Fellows program (Grant No. 2018.0460) and the project Dynamic Quantum Matter (Grant No. 2019.0068), as well as the G\"oran Gustafsson Foundation for Research in Natural Sciences and Medicine.

\appendix

\section{Liouvillian spectrum of 1D SSH model at odd lengths}
\label{app:liou}
Here we present the exact solutions to the Lindblad master equation for the SSH model with loss on the $B$ sublattice at arbitrary odd number of sites using third quantization. 
The exactly solvable dissipationless boundary mode together with the damping bulk modes give us full access to the quantum dynamics of the relaxation process, as well as a non-trivial non-equilibrium steady state (NESS). 

\subsection{Derivation of the damping matrix in third quantization}
We start by writing the Hamiltonian of an SSH chain of spinless fermions with an odd number of sites:
$\mathcal{H}_{S} = \sum_{j=1}^{N-1} t_1 a_{j,A}^\dagger  a_{j,B}+t_2 a_{j+1,A}^\dagger  a_{j,B}+ \text{H.c.}$. 
The operator $a_{j,A}^\dagger (a_{j,B})$ creates (annihilates) a fermion on the sublattice A(B) in the $j$-th unit cell and they satisfy fermionic anticommutation relations: $\{a_{j,\alpha}, a_{j',\alpha'}^\dagger\}=\delta_{j,j'} \delta_{\alpha,\alpha'}$. 
The last unit cell is broken and the total number of sites becomes odd, $L = 2N-1$.
Adding particle loss on the $B$ sublattice with the jump operator $\hat{L}^l_{j,B}=\sqrt{\gamma^l_{0,B}} a_{j,B}$ yields a quadratic Lindbladian from the Lindblad master equation in Eq.~(\ref{eq:lin}), of which the rapidity spectrum can be obtained via third quantization \cite{prosen2008, prosen2010ex, prosen2010sp}. 
We briefly review the approach and adopt the same Majorana representation as in Ref.~\cite{emil2022l}. 
One spinless fermion can be mapped to two Majorana fermions per site: 
\begin{gather}
a_{j,A}=\frac{1}{2}(c_{j,A} - id_{j,A}), \ 
a_{j,B}=\frac{1}{2}(d_{j,B} + ic_{j,B}). \label{eq:mj}
\end{gather} 
This particular choice of mapping decouples two sectors in the damping matrix belonging to different Majorana fermions species $c$ and $d$. 
Let us group them into a whole set under a vector notation $\u{w}=(w_1, w_2, \dots, w_{2L})^T = (c_1, \dots, c_L, d_1, \dots, d_L)^T$. 
Majorana fermions are their own {anti-particles} as evinced by $w_j^\dagger = w_j$, and they obey anticommutation relations $\{ w_{j}, w_{k} \} = 2\delta_{j,k}$ such that each Majorana {fermion} squares to one.
The density matrix $\rho$ in the original Hilbert space of dimension $2^L \times 2^L$ is now embedded in the set of Majorana operators with the new form $P_{\u{\alpha}}= w_1^{\alpha_1}w_2^{\alpha_2}\cdots w_{2L}^{\alpha_{2L}}$, where $\alpha_{j} \in \{0, 1\}$. 
In the Majorana representation, the Hamiltonian and Lindblad dissipators take the form:
$\mathcal{H} = \sum_{j,k} w_j H_{j,k} w_k$ and 
     $M_{ij} = \sum_{\nu = g, l} (l^\nu_{i,\mu})^T  (l^\nu_{\mu, j})^*$. More precisely,
  \begin{align}
    \u{w}^T H \u{w} &= 
    \begin{pmatrix}
      \u{c}^T & \u{d}^T
    \end{pmatrix}               
    \begin{pmatrix}
              H_0 & 0 \\
              0 & H_0
          \end{pmatrix}     
          \begin{pmatrix}
          \u{c} \\ \u{d}
    \end{pmatrix}, \notag \\  
     \u{w}^T M \u{w} &=
    \begin{pmatrix}
      \u{c}^T & \u{d}^T
    \end{pmatrix}               
    \begin{pmatrix}
             M_1 & iM_2 \\
               -iM_2 & M_1
          \end{pmatrix}     
          \begin{pmatrix}
          \u{c} \\ \u{d}
    \end{pmatrix},  \label{eq:hm}
       \end{align} 
where
      \begin{align}
      H_0 &= \frac{i}{4}
          \begin{pmatrix}
           0 & t_1 & & & & &   \\
           -t_1 & 0 & -t_2 & & &  &  \\
            & t_2 & 0 & t_1 &  & &   \\
                   &    & &  \ddots &  & &  \\
        & &   & &  &   0 & -t_2 \\
    &   & &  & &     t_2 & 0   
         \end{pmatrix}, \notag \\
               M_1 &= \frac{1}{2} 
          \begin{pmatrix}
           \gamma_A  & & & & & &   \\
          & \gamma_B &  & & & & & \\
            &  & \gamma_A &  & & & \\        
            & & & &  \ddots &  &  \\
     &  &    & & &    \gamma_B &  \\
     &   &  & & & &  \gamma_A    
         \end{pmatrix},  \notag \\
       M_2 &= M_1(\gamma_\alpha \rightarrow \eta_\alpha), \quad \alpha = A, B,
         \label{eq:hb}
         \end{align}
where $\gamma_\alpha$'s and $\eta_\alpha$'s stand for the sum
    and the imbalance of the most generic loss and gain dissipators on two sublattices
    \begin{gather}
         2\gamma_{\alpha} =  |\gamma_{0,\alpha}^l| + |\gamma^g_{0,\alpha}|, \quad
         2\eta_{\alpha} =  |\gamma_{0,\alpha}^l| - |\gamma^g_{0,\alpha}|, \notag \\
         \text{ for } \hat{L}^l_{j,\alpha}=\sqrt{\gamma^l_{0,\alpha}} a_{j,\alpha}, \ \hat{L}^g_{j,\alpha}=\sqrt{\gamma^g_{0,\alpha}} a_{j,\alpha}^\dagger, \ \forall j.
         \label{eq:eg}
    \end{gather}
Over the $2^{2L}$-dimensional Liouville space $\mathcal{K}$, we now define adjoint fermionic operators that {annihilate and create} Majorana fermions: $\varphi_j |P_{\underline{\alpha}} \rangle = \delta_{\alpha_j, 1} | w_j P_{\underline{\alpha}} \rangle$, $\varphi_j^\dagger |P_{\underline{\alpha}} \rangle = \delta_{\alpha_j, 0} | w_j P_{\underline{\alpha}} \rangle$, which obey $\{\varphi_j, \varphi_k^\dagger \} = \delta_{j,k}$. 
Since the Fermi parity $\mathcal{P}_F = (-1)^{\sum_j \varphi_j^\dagger \varphi_j}$ is conserved, $[\hat{\mathcal{L}}, \mathcal{P}_F ] = 0$ and $ (\mathcal{P}_F)^2 = 1$, we can represent the Liouvillian in the even parity sector $ \mathcal{P}_F = + 1$ as
   \begin{gather}
     \hat{\mathcal{L}}_+ = \frac{1}{2} 
       \begin{pmatrix}
           \u{\varphi}^\dagger \cdot & \u{\varphi} \cdot
       \end{pmatrix} 
       \begin{pmatrix}
         -X^\dagger & iY \\
         0 & X
       \end{pmatrix}
         \begin{pmatrix}
           \u{\varphi} \\
          \u{\varphi}^\dagger
       \end{pmatrix}
       - A_0, \label{eq:lplus}
    \end{gather}
with $X = -4i H + M + M^T$, $Y = -2i(M - M^T)$, and $A_0 = \frac{1}{2}\tr[X]$. In the chosen Majorana fermions representation in Eq.~(\ref{eq:mj}), the full damping matrix is decoupled in different Majorana fermions species. The diagonal and off-diagonal blocks read
  \begin{gather}
     X = \begin{pmatrix}
             X_c & 0 \\
             0 & X_d
            \end{pmatrix}, \quad
     Y = 4 \begin{pmatrix}
              0 &  M_2 \\
               - M_2 &  0 
          \end{pmatrix}, \label{eq:xyo}
       \end{gather}
with $X_c = X_d = -4iH_0 + 2M_1$. 
The upper-triangular form of $\hat{\mathcal{L}}_+$ indicates that the eigenvalues of the Liouvillian coincide with those of the damping matrix $X$ \cite{prosen2008,emil2022l}. 
After proper diagonalization, we are able to express the Liouvillian in terms of {\it rapidities} $\beta$ and {\it normal master modes} {(NMMs)} $b'$, $b$:
    \begin{gather}
      \hat{\mathcal{L}}_+ = - \sum_{m=1}^{L} \beta_m (b'_{c,m} b_{c,m} + b'_{d,m} b_{d,m} ). \label{eq:nmm}
   \end{gather}
Here, $\beta_m = \beta_{c,m} = \beta_{d,m}$ with the band index $m$ and {NMMs satisfy} the anticommutation relations $\{ b'_{c,m}, b_{c,l} \}=\{ b'_{d,m}, b_{d,l} \} = \delta_{m,l}$.

Furthermore, we can map the damping matrix in the Liouvillian to the SSH Hamiltonian with additional non-Hermitian imbalanced chemical potential on the two sublattices. 
For the (uniform) loss on the B sublattice only, we arrive at
 \begin{gather}
   X_c = X_d = \frac{\gamma}{2}\times \mathbbm{1}_{L \times L} + iU\mathcal{H}_{\text{NH}}U^{-1}, \quad \mathcal{H}_{\text{NH}} = \mathcal{H}_S+ i \Upsilon, \notag \\
          \mathcal{H}_S = 
            \begin{pmatrix}
           0 & t_1  & & & &   \\
           t_1  & 0 & t_2  & & &    \\
            & t_2  & 0 &  & &   \\
                     & & & \ddots & & \\     
           & & &  & 0 & t_2 \\
           & & & &  t_2 & 0   
     \end{pmatrix}_{L \times L}, \notag \\
       \Upsilon = \frac{\gamma}{2} \begin{pmatrix}
      1 & & & & & \\
      & -1 & & &  & \\
      & & 1 & & & \\
      & & & \ddots & & \\
      & & & & -1 & \\
      & & & &  & 1
    \end{pmatrix}_{L \times L}, \label{eq:dma}
 \end{gather}
with the unitary transformation $U = \text{diag}\{1,i,1,i,\dots,i,1\}$ and $\gamma = |\gamma_{0,B}^l|/2$.
Therefore, the rapidity spectrum satisfies the relation
   \begin{gather}
     \beta_m = \frac{\gamma}{2} + i E_m, \label{eq:b_e}
   \end{gather}
where $E_m$ denotes the eigenvalues of the matrix $\mathcal{H}_{\text{NH}} = \mathcal{H}_S+ i \Upsilon$.

\subsection{Boundary mode and full rapidity spectrum with spectral mirror symmetry}
\begin{figure}[t]
\centering
\includegraphics[width=1\linewidth]{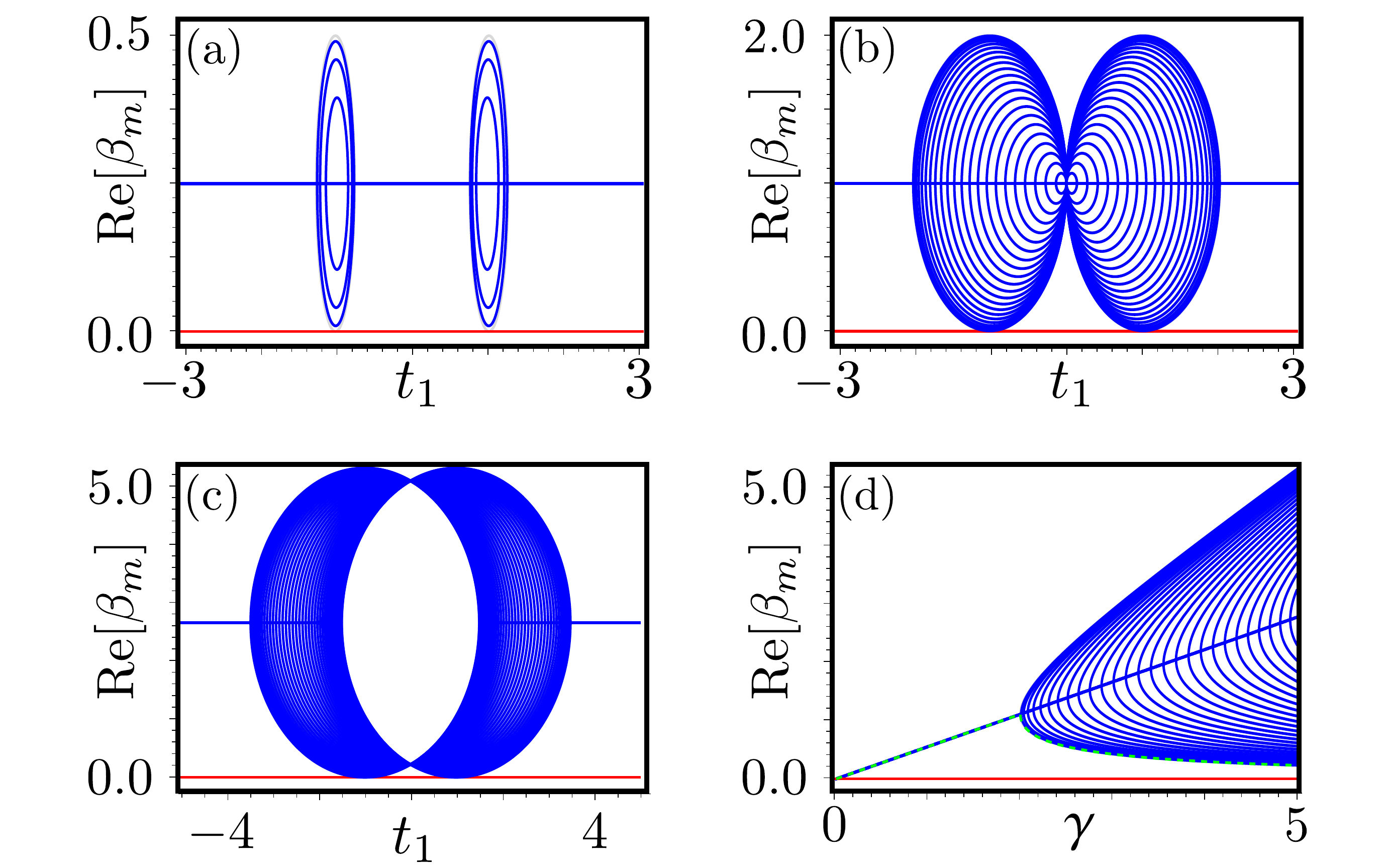}  
\caption{(a),(b) Real part of the rapidity spectrum as a function of $t_1$ for $ t_2 = 1$ and $N = 46$. The gray lines show the structure of the periodic system while the blue and red lines correspond to the bulk and {boundary} modes under OBC. We set different loss dissipation strengths on the B sublattice: (a) $\gamma = 0.5$, (b) $\gamma = 2$, (c) $\gamma = 5$. (d) Real part of the rapidity spectrum as a function of $\gamma$ for $t_1 = 2$, $t_2 =1$, and $N = 46$. The green line denotes the half of the {bulk dissipative} gap estimated according to \Cref{eq:gap} for $N \gg 1$.
}
\label{fig:beta}
\end{figure}
Under open boundary conditions (OBC), for the SSH chain with an odd number of sites $ L = 2N-1$, we first identify the zero-rapidity boundary mode 
     $\beta_{m=0}= 0$.

It comes from the zero-energy boundary state of the SSH Hamitonian:
  \begin{gather}
    \mathcal{H}_S \tilde{\u{\psi}}_{m} = \tilde{E}_m \tilde{\u{\psi}}_{m},  \quad
     \tilde{E}_0 = 0, \  \tilde{\u{\psi}}_{0} = \mathcal{N}  \begin{pmatrix} r \\ 0 \\ r^2  \\ 
     \cdots \\ 0 \\ r^N \end{pmatrix}, \  r = -\frac{t_1}{t_2}, \label{eq:edge}
  \end{gather}
where the normalization factor reads $\mathcal{N}^2=(1-r^2)/[r^2(1-r^{2N})]$.
  It is straightforward to check that
     \begin{gather}
    (H_{\text{S}} + i \Upsilon) \tilde{\u{\psi}}_{R0} = E_0 \tilde{\u{\psi}}_{R0}, \quad E_0 = i\frac{\gamma}{2},
     \end{gather}
which leads to zero rapidity from \Cref{eq:b_e},
 \begin{gather}
   \beta_0 = 0, \quad \u{\psi}_{R0} = \u{\psi}_{L0} = \u{\tilde{\psi}}_0. \label{eq:beta0}
 \end{gather} 
Physically, this is consistent with the observation that the wave function of the zero-energy {boundary} mode is fully suppressed on the B sublattice for an SSH chain with an odd number of sites.
The {boundary} mode survives under the dissipation on the frustrated sublattice in the infinitely long-time limit.

For a chain with an odd number of sites, the bulk spectrum fulfils a spectral mirror symmetry:  $\beta_{\pm}^{\text{OBC}}(q) =  \beta_{\pm}^{\text{OBC}}(-q)$. 
The mirror symmetry allows one to establish a rigorous relation between the spectrum under OBC and the spectrum under periodic boundary conditions (PBC)\cite{flore2019e,elisabet2020,elisabet2022},
   \begin{align}
 \beta_{\pm}^{\text{OBC}}(q) &=  \beta_{\pm}^{\text{PBC}}(q)  \notag
 \\ &= \frac{\gamma}{2} \pm i \sqrt{t_1^2 + t_2^2 + 2t_1t_2 \cos(q) - \frac{\gamma^2}{4}}, 
   \end{align}
with $q = \frac{\pi m'}{N}, \quad m' = 1, 2, \dots, N-1$. 
We plot the real part of the rapidity spectrum as a function of the hopping amplitude and the dissipation strength in \Cref{fig:beta}.
 
From the exact solutions, we can also obtain the {dissipative} gap that separates the {boundary} and bulk modes, together with the relaxation time $\tau$ for the system to evolve to the {boundary} mode in the long chain limit $N \gg 1$,
   \begin{gather}
     {\Delta_{\text{bulk}}} = 2\min \{ \re[\beta_{m\ne 0 }]\}, \quad 
      \tau \sim \frac{1}{\Delta}, 
   \end{gather}
   where 
   \begin{gather}
     {\Delta_{\text{bulk}}} = 
     \begin{cases}
       \gamma, & ||t_1| - |t_2|| \ge \gamma/2 ; \\
       \gamma - \sqrt{\gamma^2 - 4 (|t_1| - |t_2|)^2}, &  ||t_1| - |t_2|| < \gamma/2.
     \end{cases}        \label{eq:gap}
   \end{gather}

\subsection{Exact eigenmodes of the damping matrix and dynamical observables}
The complete set of eigenvectors of the damping matrix can be used to obtain dynamical observables. 
Apart from the exact boundary mode in Eq.~(\ref{eq:edge}), next we show how to construct the analytical solutions to the bulk eigenmodes.

To make the solutions more generic, we start from a Bloch form of the non-Hermitian tight-binding Hamiltonian {$\mathcal{H}_\text{NH}$ into which the damping matrix can be transformed through} Eq.~(\ref{eq:dma}),
\begin{gather}
  \mathcal{H}_{\text{NH}}(q) = \vec{h}(q) \cdot \vec{\sigma}, \quad \vec{h} = (h_x, h_y, h_z), \quad \vec{\sigma} = (\sigma_x, \sigma_y, \sigma_z), \label{eq:nhh}
\end{gather}
where $h_{x}, h_{y} \in \mathbb{R}$ and  $h_{z} \in \mathbb{C}$. The non-Hermitian term enters into the sublattice potential. From the eigenvalue equations, 
$\mathcal{H}_{\text{NH}}\u{u}_R(q) = E(q) \u{u}_R(q)$ and $\mathcal{H}_{\text{NH}}^\dagger \u{u}_L(q) = E^*(q) \u{u}_L(q)$, one obtains the right and left eigenmodes {of} the Bloch Hamiltonian,
 \begin{align}
    \u{u}_{R,\pm}(q) &= \frac{1}{\sqrt{2h(h{\mp} h_z)}}
   \begin{pmatrix}
   h_x - ih_y \\
   \pm h - h_z
    \end{pmatrix}, \notag \\
    \u{u}^*_{L,\pm}(q) &= \frac{1}{\sqrt{2h(h{\mp} h_z)}} \begin{pmatrix}
   h_x + ih_y \\
   \pm h - h_z
    \end{pmatrix}.
 \end{align}
{Due to the spectral mirror symmetry, we establish the relation for energy eigenvalues switching from PBC to OBC,}
\begin{gather}
  E^{\text{OBC}}_{\pm}(q) =  E^{\text{PBC}}_{\pm}(q)  =  \pm \sqrt{h_x^2 +h_y^2 + h_z^2},
\end{gather}
{with $q = \frac{\pi m'}{N}, \quad m' = 1, 2, \dots, N-1$. 
Therefore, the bulk eigenmodes under OBC can be constructed as a superposition of the PBC eigenmodes at opposite momenta,}
  \begin{gather}
    \u{\tilde{\psi}}_{R/L, \nu}(q, j) = \frac{1}{\sqrt{2N}} \left( e^{iqj}  \u{u}_{R/L, \nu}(q) - e^{-iqj} \u{u}_{R/L, \nu}(-q) \right).
    \label{eq:bulk}
  \end{gather}
{Here, the minus sign is determined by the boundary condition, ensuring that the overall wavefunction vanishes on the B site of the last broken unit cell.} 

If we denote the band index as $m \in \{0, (\pm, q) \}$, {it is straightforward to check that the boundary and bulk modes} in Eqs.~(\ref{eq:edge}) and (\ref{eq:bulk}) satisfy the biorthogonal relations \cite{brody2013, flore2018, elisabet2020}:
    $\u{\tilde{\psi}}^*_{L, m} \cdot  \u{\tilde{\psi}}_{R, l} = \delta_{m,l}$.
The eigenmodes of the damping matrix can be obtained with an additional unitary transformation {according to} Eq.~(\ref{eq:dma}), 
  \begin{gather}
     \u{\psi}_{Rm}  = U  \ \tilde{\u{\psi}}_{Rm}, \quad   \u{\psi}_{Lm}  = U  \ \tilde{\u{\psi}}_{Lm}, \label{eq:cb}
    \end{gather}
and they inherit the biorthogonal relations:
  \begin{gather}
     {\u{\psi}}_{Lm}^* \cdot {\u{\psi}}_{Rl} = \tilde{\u{\psi}}_{Lm}^* \cdot \tilde{\u{\psi}}_{Rl}  = \delta_{m,l}. \label{eq:bio}
  \end{gather} 
For the SSH model with loss on the {B sublattice}, we obtain the bulk modes by identifying
   \begin{gather}
     h_x = t_1 + t_2 \cos(q), \quad h_y = t_2 \sin(q), \quad h_z = i \frac{\gamma}{2}.
   \end{gather}

It is then convenient to {resolve} the time evolution {of} the observables {with} the complete set of {exactly} solvable eigenstates of the damping matrix.
Applying the anticommutation relations of Majorana fermions $\{w_j, w_k\} = 2\delta_{j,k}$ to the Lindblad master equation {in \Cref{eq:lin}}, we {arrive at} the equation of motion for the covariance matrix $C_{jk}(t) = -\tr [w_jw_k\rho(t)]+\delta_{j,k}$:
  \begin{gather}
    \partial_t {C}(t) = -{C}(t)X - X^\dagger {C}(t) + iY. \label{eq:eom1}
  \end{gather}
For the trivial steady state, 
  \begin{gather}
 \partial_t C_s = 0,  \quad  C_s  = i \begin{pmatrix}
    0 &  \mathbbm{1}_{L \times L}  \\
    - \mathbbm{1}_{L \times L} & 0
     \end{pmatrix}.
   \end{gather}
Now, we define the {expectation value of a local observable with respect to} the trivial NESS: $\tilde{C}(t) = C(t)-C_s$.
Starting from an arbitrary initial configuration that is not trivial $\tilde{C}(0) \ne 0$, we can integrate the {equation of motion} and implement the diagonalized damping matrix in the exponential,
  \begin{gather}
   X = \sum_m \sum_{\mu = c,d} \beta_{m} |\Theta^\mu_{Rm}\rangle \langle \Theta^\mu_{Lm}|,  \\
      |\Theta^c_{R(L)m}\rangle = \begin{pmatrix}
                            \u{\psi}_{R(L)m} \\
                          0
                     \end{pmatrix}, \quad
             |\Theta^d_{R(L)m}\rangle = \begin{pmatrix}
                    0 \\
                   \u{\psi}_{R(L)m} 
                 \end{pmatrix}.           \notag   
                 \end{gather}
{Due to} the biorthogonality of the basis, {one reaches a compact form,}
 \begin{align}
    & \tilde{C}(t)  \notag \\
    &= \sum_{m,m'} \sum_{\mu, \mu'} e^{-(\beta_m + \beta^*_{m'})t} |\Theta^{\mu' }_{Lm'} \rangle \langle \Theta^{\mu' }_{Rm'} | \tilde{C}(0) |\Theta^\mu_{Rm} \rangle \langle \Theta^\mu_{Lm} |. \label{eq:pmf}
 \end{align}
At $t=0$, we choose the system to be a completely filled chain: $\langle n_j (0)\rangle = 1, \forall j$. {It corresponds to} a covariance matrix,
  \begin{gather}
    \tilde{C}(0) = -2i 
     \begin{pmatrix}
    0 &  \mathbbm{1}_{n \times n}  \\
    - \mathbbm{1}_{n \times n} & 0
     \end{pmatrix},  \label{eq:ini}
  \end{gather}
{that} selects $\mu \ne \mu'$. 
{Going} back to the physical {space consisting of spinless fermions, we} define the single-particle correlator $Q_{jk} (t) =  \tr [a_j^\dagger a_k \rho(t) ]$. 
The mapping to Majorana fermions in Eq.(~\ref{eq:mj}) yields
  \begin{gather}
   Q_{jk}(t) =
          \frac{i}{4} \sigma(j,k) \left[ C_{j,{k+L}} (t) + C_{k,{j+L}} (t) \right], \label{eq:qq}
  \end{gather}
{where} the phase factor depends on whether or not the correlation resides on the same sublattice,
  \begin{gather}
    \sigma(j,k) = \begin{cases}
                 1, & j+k = \text{even}; \\
                 (-1)^{j}\cdot(-i), & j+k = \text{odd}.
               \end{cases}
  \end{gather}
Combined with \Cref{eq:pmf,eq:ini}, the {single-particle} correlator takes an explicit form in terms of the exact solutions of the damping matrix,
 \begin{align}
    Q_{jk} (t)  =& \sigma(j,k) \sum_{m,m'} \sum_{l =1}^{L} e^{-(\beta_m + \beta^*_{m'})t} \notag \\  
 & \times  \psi_{Lm}^* (j) \psi_{Lm'}(k) \cdot \psi_{Rm} (l) \psi^*_{Rm'} (l). \label{eq:spc}
  \end{align}
The particle number operator then reads
\begin{align}
  & \langle \tilde{n}_j(t) \rangle = \langle n_j(t)\rangle - \langle n_{j} \rangle_s \label{eq:pn} \\
    &=  \sum_{m,m'} \sum_{l =1}^{L} e^{-(\beta_m + \beta^*_{m'})t} \psi_{Lm'} (j)  \psi^*_{Rm'} (l) \psi_{Rm} (l) \psi_{Lm}^* (j), \notag
\end{align}
where  $\langle n_{j} \rangle_s = 0$ denotes the trivial steady state.

\section{Non-equilibrium steady state for generic loss and gain}
\label{app:ssc}

Below, we show the analytic structure of the non-trivial NESS with generic loss on the B sublattice and gain on the A sublattice. It also leads to a closed form for the steady state current.

{Given} the set of Lindblad dissipators $\hat{L}^l_{j,B}=\sqrt{\gamma^l_{0,B}} a_{j,B}, \hat{L}^g_{j,A}=\sqrt{\gamma^g_{0,A}} a_{j,A}^\dagger, \forall j $, the damping matrix {from Eqs.~(\ref{eq:hm})--(\ref{eq:xyo}) becomes}
 \begin{gather}
   X_c = X_d = \frac{(\gamma_A + \gamma_B)}{2}\times \mathbbm{1}_{L \times L} + iU(H_{\text{S}}+ i \Upsilon)U^{-1}, \notag \\
       \Upsilon = \frac{(\gamma_B - \gamma_A)}{2}  \begin{pmatrix}
      1 & & & & & \\
      & -1 & & &  & \\
      & & 1 & & & \\
      & & & \ddots & & \\
      & & & & -1 & \\
      & & & &  & 1
    \end{pmatrix}_{L \times L},
 \end{gather}
 {with $\gamma_A = |\gamma_{0,A}^g|/2$ and $\gamma_B = |\gamma_{0,B}^l|/2$.} The Bloch Hamiltonian of $\mathcal{H}_{\text{NH}} = \mathcal{H}_S+ i \Upsilon = \vec{h}(q) \cdot \vec{\sigma}$ reads 
    \begin{gather}
     h_x = t_1 + t_2 \cos(q), \quad h_y = t_2 \sin(q), \quad h_z = i \frac{(\gamma_B - \gamma_A)}{2}.
   \end{gather}
It {encompasses} all the ingredients {to} solving the rapidity spectrum, 
  \begin{align}
  \beta_0 &= \gamma_A;  \\
    \beta_{\pm}(q) &= \frac{\gamma_A + \gamma_B}{2} \notag \\
    & \pm i \sqrt{t_1^2 + t_2^2 + 2t_1t_2 \cos(q) - \frac{(\gamma_B - \gamma_A)^2}{4}}, \notag
    \end{align}
{together with} the eigenmodes according to Eqs.~(\ref{eq:nhh})--(\ref{eq:cb}).
   
{To find the NESS that} satisfies 
\begin{gather}
 \partial_t C_s = 0, \quad X^\dagger C_s + C_sX = iY,
\end{gather}
{we begin} by rewriting each matrix into the blocks,
 \begin{gather}
    X = \begin{pmatrix}
   X_c & 0 \\
   0 & X_d
   \end{pmatrix}, \quad
   C_s = \begin{pmatrix}
   0 & D_s \\
   -D_s & 0
   \end{pmatrix}, \notag \\
   Y = \begin{pmatrix}
   0 & Y_A + Y_B \\
   -(Y_A + Y_B) & 0
   \end{pmatrix}.
 \end{gather}
{It simplifies to}
\begin{gather}
  X_c^\dagger D_s + D_sX_c = i(Y_A + Y_B), \notag \\
 Y_A = 
  \begin{pmatrix}
   -2\gamma_A & & & & & \\
   & 0 & & & & \\
   & & -2\gamma_A & & & \\
  & &  & \ddots  &  & \\
  & & & & 0 & \\
   & & & & & -2\gamma_A
  \end{pmatrix}, \notag \\
  Y_B = 
  \begin{pmatrix}
   0 & & & & & \\
   & 2\gamma_B & & & &  \\
   & & 0 & & &  \\
  & &  & \ddots  & &  \\
  & & & & 2\gamma_B & \\
 &  & & & & 0 
  \end{pmatrix}.
  \end{gather}
{In the presence of generic loss and gain, an analytical solution to the steady state covariance matrix follows}
\begin{align}
    D_s &= i (\mathbbm{1}_{L \times L} + \Delta D),  \label{eq:Ds} \\
    \Delta D &= \sum_{m,m' \ne 0} p_{m,m'} |\u{\psi}_{Lm} \rangle \langle \u{\psi}_{Lm'}| + c_0|\u{\psi}_{L0} \rangle \langle \u{\psi}_{L0}|.
   \notag
\end{align}
{While} the matrix $\mathbbm{1}_{L\times L}$ in $D_s$ gives back the state of an empty chain {due to loss, $c_0$ denotes the contribution from the boundary mode immune to loss and coefficients $p_{m,m'}$ denote the overlaps between the bulk modes arising from gain,}
\begin{gather}
c_0 = p_{0,0} = -2, \quad p_{m,m'} = \frac{\langle \psi_{Rm}|2Y_A |\psi_{Rm'} \rangle}{\beta_m^* + \beta_{m'}}. 
\end{gather}
{We consider the scenario when $\gamma_B \gg \gamma_A$ and $\gamma_A \to 0^{+}$. The} non-trivial NESS can be connected to a sum of the empty state and the boundary mode. {Since the former gives zero particle number occupation, regardless of the initial conditions, the dissipative system always  relaxes to a unique NESS with a localization structure approximating  that of the boundary mode.}

For the {measurement} of the steady-state current $J_S = (i/L) \sum_j (\langle a_j^\dagger a_{j+1} \rangle_s - \langle a_{j+1}^\dagger a_j \rangle_s )$, we reach a closed form {by substituting Eqs.~(\ref{eq:qq}) and (\ref{eq:Ds}) to the above definition:}
\begin{gather}
  J_{S} = -\frac{1}{L} \sum_{j}^{L-1} \sum_{m,m'\ne 0} p_{m,m'} \psi_{Lm}(j) \psi_{Lm'}^* (j+1). \label{eq:jss}
\end{gather}

\section{Liouvillian spectrum of 2D Chern insulator}
\label{app:chern}
\begin{figure}[t]
\centering
     \includegraphics[width=0.8\linewidth]{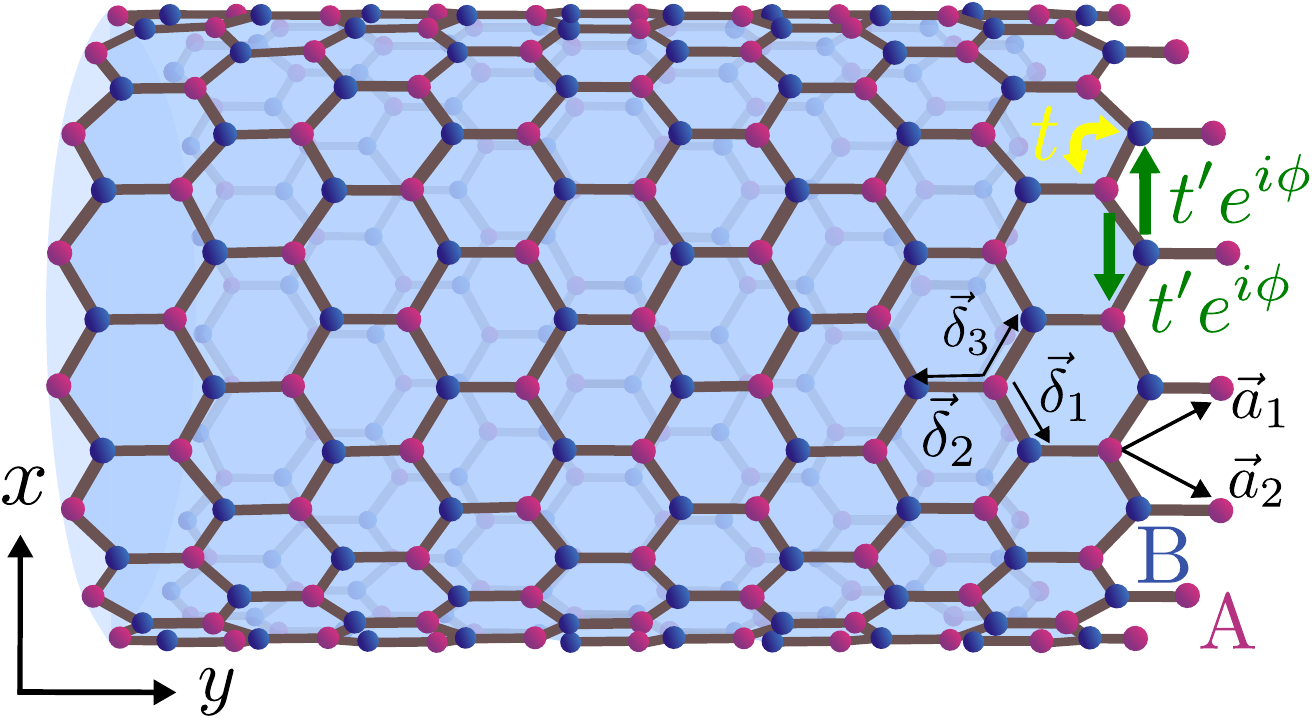}  
\caption{Illustration of a Chern insulator on a honeycomb lattice consisting of two sublattices A and B. 
We consider a system with PBC in the $x$ direction and OBC in the $y$ direction, yielding a cylinder geometry. 
We wrap the cylinder such that its edges are of zigzag type  on the left and bearded type on the right. 
The vectors $\vec{a}_1$ and $\vec{a}_2$ indicate the Bravais lattice vectors of the honeycomb lattice, while $\vec{\delta}_1$ indicate the vectors in the three possible bond directions. 
The sites are coupled by nearest-neighbor hopping $t$ (depicted in yellow) and next-nearest-neighbor hopping $t'$ (depicted in green). 
We further require that the next-nearest-neighbor hoppings along the $x$ direction acquire an additional phase factor $e^{i \phi}$ to open a Chern/Haldane insulator gap. In the text, we adopt $\phi=\pi/2$.}
\label{fig:lattice}
\end{figure}
Here, we present more details on the {derivation of the exactly} solvable rapidity spectrum {of} a 2D dissipative Chern insulator \cite{haldane1988, flore2019e} through a mapping to the generic SSH model.

As shown in Fig.~\ref{fig:lattice}, we place the honeycomb lattice on a cylinder and adopt PBC along the $x$ direction with $M$ unit cells and OBC along the $y$ direction with $N$ unit cells. The last unit cell along the $y$ direction is broken, leading to $L_y = 2N-1$ sites.
We choose the zigzag edge on the left boundary and the bearded edge on the right boundary. The nearest-neighbor and next-nearest-neighbor hopping {strengths are denoted by real parameters} $t$ and $t'$. 
The complex hoppings $t'e^{i\phi}$ with $\phi=\pi/2$ are allowed only within unit cells along the $y$ direction. 
{Setting the honeycomb lattice constant to $1$, different} unit cells are connected by the vectors
\begin{gather}
  {\vec{a}_1 = (\frac{\sqrt{3}}{2}, \frac{3}{2}), \quad 
  \vec{a}_2 = (-\frac{\sqrt{3}}{2}, \frac{3}{2}).}
\end{gather}
First, we {perform} a Fourier transform on the $x$ components {of the fermionic annihilation and creation operators ($\alpha = A, B$),} 
\begin{gather}
\vec{r} = (l,j), \quad
  a_\alpha (l,j) = \frac{1}{\sqrt{M}} \sum_{k_x} e^{ik_x l} a_\alpha (k_x,j).
\end{gather}
For each $k_x = 2\pi m'/M, m' = -M/2, \dots, M/2-1$, {the real-space Hamiltonian shares the form}
\begin{align}
  & H(k_x) |_{L_y \times L_y}  \\
  &= \begin{pmatrix}
    H_A & H_{A\leftarrow B} & & & & \\
    H_{A\leftarrow B}^\dagger & H_{B} &  H_{B\rightarrow A}^\dagger & & \\
    & H_{B\rightarrow A} & H_A &  & \\
    & & & \ddots & & \\
     & & & &  {H_B} &  {H_{B\rightarrow A}^\dagger} \\
    & & & & {H_{B\rightarrow A}} & {H_A}
  \end{pmatrix}.\notag
\end{align}
The matrix elements can be obtained as follows: 
\begin{align}
  H_A: &\sum_l t' e^{i\phi} a_A^\dagger(l,j) a_A(l+\vec{a}_1-\vec{a}_2,j) + \text{H.c.} \notag \\
  &= \sum_{k_x} -2t' \sin(k_1 - k_2)a_A^\dagger (k_x,j) a_A(k_x,j), \notag \\
    H_B: &\sum_l t' e^{-i\phi} a_B^\dagger(l,j) a_B(l+\vec{a}_1-\vec{a}_2,j) + \text{H.c.} \notag \\
    &= \sum_{k_x} 2t' \sin(k_1 - k_2)a_B^\dagger (k_x,j) a_B(k_x,j), \notag \\
  H_{A\leftarrow B}:  &\sum_l t a_A^\dagger (l,j) a_B(l,j) + t a_A^\dagger (l,j) a_B(l-(\vec{a}_1 - \vec{a}_2),j) \notag \\
  &= \sum_{k_x} t(1+e^{-i(k_1 - k_2)})a_A^\dagger (k_x,j) a_B(k_x,j), \notag \\
    H_{B\rightarrow A}: & \sum_l t a_A^\dagger (l,j+1) a_B(l-1,j) \notag \\
    &= \sum_{k_x} t e^{-i(k_1 - k_2)/2}a_A^\dagger (k_x,j+1) a_B(k_x,j),
\end{align}
{where $k_i = \vec{k} \cdot \vec{a}_i$, $i=1,2$.}

Next, let us absorb the {complex phases appearing in the} hopping terms by the {transformation}
 \begin{gather}
 a_A(k_x,j) \to e^{-i(k_1 - k_2)/2}a_A(k_x,j).
 \end{gather}
It allows us to map $H(k_x)$ to a generic SSH model with real hopping terms as before,
\begin{align}
\mathcal{H}_{S} (k_x) &= \sum_{j=1}^{N-1} t_1 a_{j,A}^\dagger  a_{j,B}+t_2 a_{j+1,A}^\dagger  a_{j,B}+ \text{H.c.} \notag \\
& + \sum_{j} \sum_{\alpha=A,B}  \epsilon_{\alpha} a_{j,\alpha}^\dagger a_{j,\alpha}  \label{eq:rox1}
\end{align}
with $t_1 =2t\cos(k_x/2),  t_2=t, \epsilon_A=-\epsilon_B=-2t'\sin(k_x)$.

For simplicity, we have adopted {the} renormalized $k_x = k_1 - k_2 = \sqrt{3} k_{x,0}$, compared with the {original} value $k_{x,0}$ on the honeycomb lattice. 

{Adding the B sublattice loss $\hat{L}^l_{j,B} =\sqrt{\gamma^l_{0,B}} a_{j,B}, \forall j$, we apply the Majorana fermions representation in \Cref{eq:mj}.
The generic SSH Hamiltonian reads}
\begin{gather}
    \u{w}^T \mathcal{H}_S \u{w} = 
    \begin{pmatrix}
      \u{c}^T & \u{d}^T
    \end{pmatrix}               
    \begin{pmatrix}
              H_0 & -iH_1 \\
              iH_1 & H_0
          \end{pmatrix}     
          \begin{pmatrix}
          \u{c} \\ \u{d}
    \end{pmatrix}, \notag \\
       H_0 = \frac{i}{4} h_0, \quad H_1 = \frac{i}{4} h', 
\end{gather}
where
   \begin{align}
       h_0 &= 
        \begin{pmatrix}
           0 & t_1  & & & &   \\
           -t_1  & 0 & -t_2  & & &    \\
            & t_2  & 0 &  & &   \\
                     & & & \ddots & & \\     
           & & &  & 0 & -t_2 \\
           & & & &  t_2 & 0   
     \end{pmatrix}_{L_y \times L_y}, \\
     h' &=  \begin{pmatrix}
      -i\epsilon_A & & & & & \\
      & -i\epsilon_B & & & & \\
      & & -i\epsilon_A & & & \\
      & & & \ddots & & \\
      & & & & -i\epsilon_B & \\
    &  & & & & -i\epsilon_A
    \end{pmatrix}_{L_y \times L_y}. \notag
    \end{align}
{And the} damping matrix takes the form
\begin{gather}
     X =
    \begin{pmatrix}
      \u{c}^T & \u{d}^T
    \end{pmatrix}               
    \begin{pmatrix}
           h_0 +   2M_1 & -ih' \\
               ih' & h_0 + 2M_1
          \end{pmatrix}     
          \begin{pmatrix}
          \u{c} \\ \u{d}
    \end{pmatrix}, \notag \\
    {M_1 = \text{diag} \{0, \frac{\gamma}{2}, 0, \dots, \frac{\gamma}{2}, 0 \},}
       \end{gather}
{with $\gamma = \gamma_B = |\gamma_{0,B}^l|/2$.
It is convenient to introduce a new set of Pauli matrices $\vec{\tau}$ acting on different subspaces of} Majorana fermions species $c$ and $d$, {in order to decouple them,}
       \begin{gather}
         X = U_0 \tilde{X}U_0^{-1}, \label{eq:rox2} \\
         \tilde{X} = \begin{pmatrix}
           h_0 + h' + 2M_1 & 0 \\
           0 & h_0 - h' + 2M_1 
         \end{pmatrix} = \begin{pmatrix}
             \tilde{X}_c & 0 \\
           0 &    \tilde{X}_d
         \end{pmatrix}, \notag
       \end{gather} 
{where $U_0 = (1+i\tau^x)/\sqrt{2}$.}
The eigenmodes of $\tilde{X}_{c(d)}$ can be exactly solved by the method we have developed earlier {in Eqs.~(\ref{eq:dma})--(\ref{eq:cb}), which requires the identification of their NH Bloch Hamiltonians in \Cref{eq:nhh}. From the mapping of the damping matrix in \Cref{eq:dma}, one arrives at}
   \begin{gather}
     h_x = t_1 + t_2 \cos({k_y}), \quad h_y = t_2 \sin({k_y}), \quad h_z = i\frac{\gamma}{2} \mp \mu \epsilon_A,
   \end{gather}
   where $\mu = \pm 1$ for $c$, $d$, respectively.
   For the chiral {edge} mode, 
   \begin{gather}
   \beta_0^\mu=-i\mu\epsilon_A,\quad \u{\psi}_{R0} = \u{\psi}_{L0}  = \mathcal{N}  \begin{pmatrix} r \\ 0 \\ r^2  \\ 
     \cdots \\ 0 \\ r^N \end{pmatrix}, \notag \\
     r(k_x) = -\frac{t_1}{t_2} = - 2\cos(k_x/2),
   \end{gather}
while for the bulk modes ($k_y=\pi m'/N$ with $m'=1,\dots,N-1$), the spectral mirror symmetry $\beta_\pm (k_x,k_y) = \beta_\pm (k_x,-k_y)$ leads to
   \begin{align}
&\beta_{{\mu},\pm}^{\text{OBC}} (k_x, k_y) = \beta_{{\mu},\pm}^{\text{PBC}} (k_x, k_y)  \\
&= {\gamma}/{2}\pm i \sqrt{t_1^2+t_2^2+2t_1t_2\cos(k_y)+(i\gamma/2-\mu \epsilon_A)^2}. \notag
   \end{align}

Furthermore, we check that the trivial steady state remains the same as the {1D} SSH model (an empty state of spinless fermions along the $y$ direction), 
\begin{gather}
 C_s = i \begin{pmatrix}
             0 & \mathbbm{1} \\
           -\mathbbm{1} &   0
         \end{pmatrix}.
\end{gather}
The time evolution of the particle number distribution can be resolved as well,
 \begin{align}
   \langle n_j(t)\rangle
    =& \frac{1}{2}\sum_{m,m'} \sum_{l =1}^{{L_y}} \sum_{\mu = c,d } e^{-(\beta_{\mu,m} + \beta^*_{\mu,m'})t}  \\
    & \times \psi_{L, \mu m'} (j)  \psi^*_{R,\mu m'} (l) \psi_{R,\mu m} (l) \psi_{L, \mu m}^* (j).\notag
\end{align}
Starting from a completely filled lattice driven by the B-sublattice loss, the steady-state particle number distribution becomes that of the chiral edge mode,
\begin{equation}
 \langle n_j \rangle_{s}=\frac{r^{j-1}-r^{j+1}}{1-r^{2N}} \:\: (j \: \mathrm{odd}), \  \langle n_j \rangle_{s}=0 \:\:  (j \: \mathrm{even}).
\end{equation}

\section{Robustness of the boundary state against sublattice and bond disorders}
\label{app:dis}

In the following, we discuss the robustness of the boundary state in \Cref{eq:bm} hosted by the 1D SSH model with odd sites, and generalize its localization structure in the presence of bond disorders.

Further-neighbor interactions and weak onsite-potential disorder on the A sublattice would eliminate the boundary mode from being a steady state of the Liouvillian. 
However, with loss engineering, there are other types of disorders that the boundary mode is robust against, including strong random  disorders in the B-sublattice potentials and in the $t_1$ and $t_2$ hopping terms. 
Let us include them explicitly in the model with B-sublattice loss,
\begin{gather}
\mathcal{H}_0 = \sum_{j=1}^{N-1} t_{j,1} a_{j,A}^\dagger  a_{j,B}+t_{j,2} a_{j+1,A}^\dagger  a_{j,B}+ \text{H.c.}, \notag \\
V = \sum_j \omega_{j,B} a_{j,B}^\dagger a_{j,B}, \notag \\
\hat{L}^l_{j,B} =\sqrt{\gamma^l_{0,B}} a_{j,B}, \forall j.
\end{gather}
Casting the Hamiltonian into a matrix form, it is straightforward to check that due to local destructive interference arising from disordered nearest-neighbor hopping terms, the targeted zero-energy boundary state takes a new form,
\begin{gather}
  H_0 = \begin{pmatrix}
      0 & t_{1,1} & & & &\\
   t_{1,1} & 0 & t_{1,2} & & & \\
   &   t_{1,2} & 0& t_{2,1}  & & \\
      & & &  \ddots &  &  \\
      & & & & 0 & t_{N-1,2}\\
      & & & & t_{N-1,2} & 0
    \end{pmatrix}, \notag \\
    H_0 \underline{\psi}_0 = E_0\underline{\psi}_0: \ 
    E_0 = 0, \ 
  \underline{\psi}_0 = \mathcal{N} (r_1,0,r_2,0,\dots, 0, r_N)^T, \notag \\
  r_1 = 1, \quad r_m = \prod_{j=1}^{m-1} \left( -\frac{t_{j,1}}{t_{j,2}}\right), \  m > 1,
\end{gather}
with a normalization factor $\mathcal{N}^{-2}= \sum_{m=1}^N r_m^2$.
It still shares a localization structure fully suppressed on the B sublattice, thus immune to both loss and on-site disorder on the same sublattice. Indeed, after a proper rotation in the subspaces of different Majorana fermions species $c$ and $d$ as has been performed in Eqs.~(\ref{eq:rox1})--(\ref{eq:rox2}),
random on-site disorder acting as a real chemical potential enters the rotated damping matrix in the diagonal terms belonging to B sites and becomes purely imaginary ($\mu = +1, -1$ for $c, d$),
\begin{align}
 \tilde{X}_{c,d} =& \begin{pmatrix}
      0 & t_{1,1} & & & &\\
   -t_{1,1} & 0 & -t_{1,2} & & & \\
   &   t_{1,2} & 0& t_{2,1}  & & \\
      & & &  \ddots &  &  \\
      & & & & 0 & -t_{N-1,2}\\
      & & & & t_{N-1,2} & 0
    \end{pmatrix} \notag \\
  &+  \begin{pmatrix}
      0 & & & & & & \\
      & -i \mu \omega_{1,B} + \gamma_B & & & & & \\
      & & 0 & & & &  \\
    &  & & & \ddots & & \\
      & &  & & & -i \mu \omega_{N-1,B}+\gamma_B & \\
    & &  & & & & 0
    \end{pmatrix}. 
\end{align}
The vanishing gap of the generalized boundary mode ensures an infinite lifetime as a steady state,
\begin{gather}
   \tilde{X}_{c,d} \underline{\psi}_0 = \beta_0\underline{\psi}_0, \quad \beta_0 = 0.
 \end{gather}

\bibliography{sample}

\end{document}